# 5-DIMENSIONAL EXTENDED SPACE MODEL.

D.Yu.Tsipenyuk[(1)], V.A.Andreev [(2)]

[(1)] General Physics Institute of the Russian Academy of Sciences,
119991 Russia, Moscow, Vavilova str.38, E-mail: tsip@kapella.gpi.ru
[(2)] Lebedev Physics Institute of the Russian Academy of Sciences,
119991, Russia, Moscow, Leninsky prosp.53, E-mail: andr@sci.lebedev.ru

**ABSTRACT**

We put forward an idea that physical phenomena have to be treated in 5-dimensional space where the fifth coordinate is the interval S. Thus, we considered the (1+4) extended space $G(T;X,Y,Z,S)$. In addition to Lorentz transformations (T;X), (T;Y), (T;Z) which are in (1+3)-dimensional Minkowski space, in the proposed (1+4)d extended space two other types of transformations exist in planes (T,S); (X,S), (Y,S), (Z,S) that converts massive particles into massless and vice versa. We also consider energy-momentum-mass 5-d space which is conjugated to the time-coordinates-interval (T;X,Y,Z,S) space. The mass in above mentioned 5-d energy-momentum-mass space corresponds to the interval S in 5-d space (T,X,Y,Z,S). Well known energy-momentum 4-vector $P(1+3)=(E/c; p_x,p_y,p_z)$ in the Minkowski (T;X,Y,Z) space is transformed to the the 5-vector - $P(1+4)=(E/c; p_x,p_y,p_z,mc)$ in the extended space $G(T;X,Y,Z,S)$ and becomes isotropic for 5-vectors of the massive as well as for the massless particles.

## 1. INTRODUCTION

We consider generalization of the Einstein's special theory of relativity (SRT) on 5-dimensional space, having the metric (+ - - - -) [1-4]. The physical basis for such generalization is that fact, that in spatial theory of relativity masses of particles are scalars and do not vary in case of the elastic interactions. However it is well known, that it is possible to consider a photon as a massless particle and to describe photon by a flat wave only in infinite empty space [5]. If the photon moving in a medium or in the limited space, for example in a resonator or wave-guide, it acquires a nonzero mass [16,17].
We offered such generalization of special theory of relativity, which takes into account processes under which a mass of a particle m also would be a variable. As a mass of a particle m we, following the recommendations of the review [6], shall understand it a rest-mass, which is a Lorentz scalar. For this purpose first of all we shall construct the extension (1 + 3) - dimensional Minkowski space $M(T;X,Y,Z)$ on (1 + 4) - dimensional extended space $G(T;X,Y,Z,S)$. This space we will name as extended space. The Extended Space Model we will sign as ESM.

Similarly it is possible to consider process of a modification of a mass and for other particles, for example, electrons assuming, that mass depends on the environmental conditions and interactions. Thus, it is represented natural to expand space of parameters describing a particle, with taking into accounts that it mass can vary during interactions.

Let's reduce simple analogy. The free particle move in straight line, therefore to describe particle behavior, it is possible to limit (1 + 1) -dimensional space that consists from the time T and direction of movement X, as the remaining coordinates Y and Z remain constants. If the particle begins to interact with other objects, so that particle can leave direct line along X direction and begin to move also in a plane (YZ), it is not enough of (1+1) space already. It is necessary to expand our consideration up to (1 + 3) -dimensional space now. Precisely as well in our case, while the mass of a particle does not vary, it is possible to limit by the Minkowski space M (1,3), but if mass begins to vary, the space M (1,3) should be expanded.

Attempts to unify of gravitation and the electromagnetism have a long history. Modern approaches to this problem ascend to F. Klein [7] who showed, that the classical Hamilton's mechanics could be represented as optics in a space of higher dimensions. Later on Th. Kaluza made an attempt to generalize the Einstein's gravitation theory in order to include electromagnetism [8]. He proposed to consider (1 + 4) -dimensional space with the metric dependent on the electromagnetic-field potentials. This idea was developed by O.Klein [9], H.Mandel [10] and V.Fock [11], the model, constructed by them, was named the Kaluza-Klein theory. The trajectory of a charged particle was shown to have the form of a geodesic line with zero length in 5-dimensional space.

Yu. Rummer in his studies on 5-optics [12] has offered to attribute to the new dimension dimensionality of action and to assume that it is periodic with period equal to the Plank's constant ℏ. Note, that the rest-mass of particles was considered to be fixed quantity in all these theories, in difference from developed 5-d extended space model (ESM) [1-4].

The subsequent development of the multidimensional theories is described in [13]. Multidimensional constructions in the theory of strings and hyperstrings [14] are the subject of an independent field search.

Close to ESM is the theory advanced in [15]. In [15] there was constructed (1 + 4) space, where a mass (substance) as fifth coordinate was offered to use. However in this model (as authors of [15] write themselves) it is impossible to construct energy-momentum tensor. In ESM this defect is absent [1-4].

Our approach essentially differs from all listed above models, though, conceptually intersects with some of them.

## 2. FORMALISM OF THE EXTENDED SPACE MODEL

As a fifth additional coordinate we use the quantity already existent in Minkowski space, namely, the interval S,

$$s^2 = (ct)^2 - x^2 - y^2 - z^2 \qquad (1)$$

This quantity is conserved in usual Lorentz transformations in Minkowski space $M(T; \vec{X}) \equiv M(T; X, Y, Z)$, but it changes under rotations in the extended space $G(T, \vec{X}, S) \equiv G(T, X, Y, Z, S)$. Thus, Minkowski space $M(T; \vec{X})$ is a cone in the extended space $G(T, \vec{X}, S)$.

In Minkowski space $M(T; \vec{X})$ to each particles are compared an energy - momentum 4-vector

$$p_4 = (\frac{E}{c}, p_x, p_y, p_z). \qquad (2)$$

In the extended space $G(T, \vec{X}, S)$ we extend it up to a 5-vector

$$p_5 = (\frac{E}{c}, p_x, p_y, p_z, mc). \qquad (3)$$

The energy, momentum and mass of free particle are known to satisfy the relation [5]

$$E^2 - c^2 p_x^2 - c^2 p_y^2 - c^2 p_z^2 - m^2 c^4 = 0. \qquad (4)$$

Relation (4) is analog of a relation (1) in space $G'(E; \vec{P}, M)$, which conjugate to the space $G(T, \vec{X}, S)$. The mass m is conjugate to the interval s. Lorentz transformations in Minkowski space change energy of a particle E and momentum p, but leave constant mass m. Transformations in extended space $G(T, \vec{X}, S)$ complementary to Lorentz transformations, change also mass of a particle.

The set of magnitudes (3) defined 5-impulse, its components are saved, if the space G(T;X,S) is invariant in an appropriate direction. In particular, fifth component has a sense of a mass m that does not vary, if the particle is moving so that particle is entire time in the field with constant density of substance or density of energy. This density of external substance (energy) can be interpreted, as a component of an external force affecting on a particle.

Thus from the physical point of view our extension of SRT takes as admissible the processes in which the rest mass of particles changes. The possibility of the existence of such processes has been discussed in the literature. For instance, if a photon enters a medium or is found in a cavity or waveguide, it can be assigned a certain nonzero mass [16,17].

In difference from a usual relativistic mechanics, now we assume, that the mass of a particle m also is a variable. It should be understood that mass of a particle varies, when particle moves in area of space having a nonzero denseness of substance. As to in such areas slow down speed of light, we shall characterize its by their magnitude n - refractive index. The parameter n links speed of light in a vacuum c with speed of light in a medium v: vn = c.

According to ESM rest mass of particles - is a variable and a photon, go ahead in the medium with the refractive index n > 1, acquires nonzero mass. The probability that a photon has nonzero mass is widely discussed both theorists, and experimenters. The review of the last outcomes is contained in [18]. Our approach differs by that in ESM mass of a particle is not constant, and is determined by external effects and processes, which affected on the particle. Formally such processes within the framework of our model are described by rotations in extended space G(1+4) [1-3]. One of the main physical problems in ESM is how to compare particular interactions with appropriated to them distribution of refractive index. In each separate case, this problem is decided in its own way.

Gravitational field is one of examples of physical objects, to which compares some refractive index. From the very beginning of origins General Theory of Relativity there was a problem of its experimental checking. One of such effects is, in particular - deviation of light in a gravitational field. This deviation can be interpreted as motion of a light beam in environment with an inhomogeneous refractive index. Thus, we can attribute some refractive index to gravitational field. We will consider further well-known gravitational effects, used for the General Theory of Relativity confirmation. We will show, that all of them can be described and in frameworks ESM. Let's mark also, that now attempts in a new way to interpret the gravitational effects are considered by other authors [19-23].

Parameter n connect speed of light in vacuum c with speed of light in the medium v = c/n. With the help of it is possible to define fifth coordinate in space G(1, 4). Thus the empty Minkowski space M(1, 3) corresponds to n = 1. In this medium light move with speed c. Light transition from medium with n=1 into medium with n > 1 is interpreted as an exit of a photon from the Minkowski space and transition of light in other subspace of space G(1, 4). Such transition can be described with the help of rotations in space G(1, 4).

In empty space in a fixed reference system there are two types of various object, with zero and nonzero masses. In the extended space G(1, 4) to that objects there are corresponds 5-vectors

$$\left(\frac{\hbar\omega}{c}; \frac{\hbar\omega}{c}; 0\right) \tag{5}$$

$$(mc; 0, mc) \tag{6}$$

For simplicity we have recorded vectors (5), (6) in (1 + 2)-dimensional space. The vector (5) describes a photon with zero mass, energy $\hbar\omega/c$ and with speed c. Vector (6) describes a fixed particle with mass m.

At hyperbolic rotations on an angle θ in the plane (TS) photon vector (5) will be transformed according to the law

$$(\frac{\hbar\omega}{c}, \frac{\hbar\omega}{c}, 0) \to (\frac{\hbar\omega}{c}\cosh\theta, \frac{\hbar\omega}{c}, \frac{\hbar\omega}{c}\sinh\theta) = \left(\frac{\hbar\omega}{c}\cdot n, \frac{\hbar\omega}{c}, \frac{\hbar\omega}{c}\sqrt{n^2-1}\right). \tag{7}$$

This transformation produces a particle with mass

$$m = \frac{\hbar\omega}{c^2}\sinh\theta = \frac{\hbar\omega}{c^2}\sqrt{n^2-1}. \tag{8}$$

Under the same rotation θ the massive vector (6) is transformed as

$$(mc, 0, mc) \to (mce^\theta, 0, mce^\theta). \tag{9}$$

Massive particle changes it's mass

$$m \to me^\theta, \quad 0 \le \theta < \infty. \tag{10}$$

and energy, but conserve momentum.

Under trigonometric rotations through angle ψ in the (XS) plane photon vector (5) will be transformed under the law

$$(\frac{\hbar\omega}{c}, \frac{\hbar\omega}{c}, 0) \to (\frac{\hbar\omega}{c}, \frac{\hbar\omega}{c}\cos\psi, \frac{\hbar\omega}{c}\sin\psi) = (\frac{\hbar\omega}{c}, \frac{\hbar\omega}{cn}, \frac{\hbar\omega}{cn}\sqrt{n^2-1}). \tag{11}$$

Thus the photon acquires mass

$$m = \frac{\hbar\omega}{c^2}\sin\psi = \frac{\hbar\omega}{c^2 n}\sqrt{n^2-1}. \tag{12}$$

and velocity

$$v = c\cdot\cos\psi = \frac{c}{n}. \tag{13}$$

Massive particle vector (6) is transformed under such rotation according to the law

$$(mc, 0, mc) \to (mc, -mc\sin\psi, mc\cos\psi) = (mc, -\frac{mc}{n}\sqrt{n^2-1}, \frac{mc}{n}). \tag{14}$$

Under such transformation energy of a particle is conserved, but it's mass and momentum are changed as

$$m \to m\cos\psi = \frac{m}{n}. \tag{15}$$

and

$$0 \to -mc\sin\psi = -\frac{mc}{n}\sqrt{n^2-1}, \tag{16}$$

The important property of the (7) and (11) transformations is that mass of a photon under these transformations can have both positive and negative sign. It follows immediately from the symmetry properties of the space G(1,4). But for massive particle, which initially had positive mass, after (9) and (14) transformations mass remains positive.

## 3. REFRACTIVE INDEX OF GRAVITATIONAL FIELD.

Let's study now a problem of refractive index n of a gravitational field. If we consider a dot mass, it's gravitational field will be described by the Schwarzchild solution. We assume, that the gravitational radius $r_g$ is small and we shall consider all effects at distances $r > r_g$.

In the literature there are two expressions for an index of refractive n, appropriate to a Schwarzchild field. One of them, we shall name it $n_1$, is used in Okun papers [24,25]

$$n_1(r) = (g_{00})^{-1} = (1 - \frac{r_g}{r})^{-1} \approx 1 + \frac{r_g}{r} = 1 + \frac{2\gamma M}{rc^2} \quad (17)$$

Formula (17) was received in the assumption that in a constant gravitational field photon's frequency remains constant, but photon's wavelength and speed v varies. Other index of refractive $n_2$ is possible to receive from the formula of an interval in a weak gravitational field [5]

$$ds^2 = (c^2 + 2\varphi)dt^2 - dr^2. \quad (18)$$

Where $\varphi$ - is a potential of a gravitational field. Supposing $dr = v\, dt$ and $ds^2 = 0$, we will receive speed of a photon in the gravitational field

$$v = c\left(1 + \frac{2\varphi}{c^2}\right)^{1/2} \approx c\left(1 + \frac{\varphi}{c^2}\right). \quad (19)$$

Here it is necessary to take into account that a potential of a gravitational field $\varphi$ is negative. For a point source of mass M we have

$$\varphi = -\frac{\gamma M}{r} \quad (20)$$

Substituting expression (20) in the formula (19), we receive

$$v \approx c\left(1 - \frac{\gamma M}{rc^2}\right). \quad (21)$$

It is possible to interpret the formula (21), as a photon penetration in medium with refractive index $n_2$

$$n_2(r) = 1 + \frac{\gamma M}{rc^2} \quad (22)$$

In the case, when the speed of a particle v is comparable to the speed of light c, in the formula (21) it is necessary to take into account relativistic correction to a rest-mass m and to record it as

$$M = m_0\left(1 + \frac{\gamma M}{rc^2} + \frac{1}{2}\frac{v^2}{c^2}\right). \quad (23)$$

Appropriate refractive index looks like

$$n_2'(r) = 1 + \frac{\gamma M}{rc^2} + \frac{1}{2}\frac{v^2}{c^2}. \quad (24)$$

Such difference in definition of gravitational field refractive indexes $n_1$ and $n_2$ is connected with follow reason: in above mentioned two cases we deal with two different objects, which differently interact with a gravitational field. In ESM to these situations there are corresponds also different rotations in extended space.

## 4. GRAVITATIONAL EFFECTS IN EXTENDED SPACE MODEL

### 4.1 Speed of escape

Speed of escape $v_2$ is that speed, which should be given to a massive body, located at the Earth surface, that the body could fly away from the Earth at an indefinitely large distance. If M - is the mass of the Earth, m - mass of a body located at the Earth surface, and R - radius of this surface. Then expression for the speed of escape is

$$v_2 = \sqrt{2gR} = \sqrt{\frac{2\gamma M}{R}}. \quad (25)$$

We will receive now formula (25) using ESM methods. Let's consider a massive particle at rest, which is at infinite large distance from the Earth. Within the framework of our model such particle is described by isotropic 5-vector of energy-momentum-mass (6). Space motion in gravitational field along an axis X can be compared to the motion in extended space G(1, 4) in plane (XS) from a point with refractive index is n = 1 to point with refractive index is n(r). Such motion is described by a rotation (14).
Here rotation angle $\psi$ expressed through refractive index n. Thus the massive particle acquires speed

$$v = c\frac{\sqrt{n^2 - 2}}{n}. \quad (26)$$

As to in this case we consider a motion of a massive body, we assume natural to use refractive index $n_2$. Assuming, that it is close to unit, i.e. that

$$1 \gg \varepsilon = \frac{\gamma M}{rc^2} \quad (27)$$

we obtain

$$v \approx c\sqrt{2\varepsilon} \quad (28)$$

In case, when r = R - radius of the Earth, the formula (28) coincides with the formula (25) and gives the speed of escape $v = v_2$.

## 4.2 Red shift

Gravitational red shift usually considered as a change of a photon frequency in the case of changing of a gravitational field, in which photon move. In particular, at decreasing of field strength photon frequency also decreases [5].

However Okun offers to assume that varies not frequency but varies wavelength of a photon, and to name this effect as red displacement [24,25]. Under our judgment both cases are possible, but they corresponds to different physical situations. These cases are described by different rotation angles in different planes of 5-d extended space expressed through refractive indexes. In a general theory of relativity the formula that describes change of light frequency is [5]

$$\omega = \frac{\omega_0}{\sqrt{g_{00}}} \approx \omega_0 \left(1 + \frac{\gamma M}{rc^2}\right). \qquad (29)$$

Here $\omega_0$ is photon frequency measured in universal time, $\omega_0$ remains constant during beam of light propagation. And $\omega$ is a frequency of the same photon that measured in its own time. This frequency is various in various points of space. Let's consider the case when a massive star emitted a photon. Near to the star at small distance r photon frequency is more, than far from the star at large r. At the infinity in the flat space, where there is no gravitational field, universal time coincides with own and $\omega_0$ is an observable photon frequency.

We will consider now the same problem from the ESM point of view. Within the framework of our model isotropic 5-vector (5) is matching to a photon located in empty space with n=1. Red shift in the frame of ESM means changing of photon energy according to changing gravitation field strength. Process of photon's motion to the point with refractive index n>1, at which the change it's frequency (and energy) happens, is described by a rotation in (TS) - plane. Under this rotation photon vector will be transformed according to the (7). Hence it is distinguished, that $\omega_0$ (photon frequency in empty space) and $\omega$ (photon frequency in a field) are connected by a ratio

$$\omega = \omega_0 \cosh\theta = \omega_0 n \qquad (30)$$

We assume that for the calculation change of photon frequency it is necessary to use refractive index $n_2$. This is due to refractive index $n_1$ was found in the supposition, that photon frequency does not vary. Substituting (22) in (30), we receive the formula

$$\omega = \omega_0 n_2 = \omega_0 \left(1 + \frac{\gamma M}{rc^2}\right) \qquad (31)$$

which coincides with (29). Thus we received in extended space model for red shift the same expression, as in general theory of relativity.

From our model point of view it is necessary to consider formula (30) only as first approximation to an exact result.

Let's estimate correction to n appropriated to ESM. When photon merged in the area with n > 1, it gains a nonzero mass. For this reason part of photon's energy can be connected not to the frequency, but with the mass.

To estimate magnitude of this energy we will consider follow mental experiment - measurement of the photon frequency, when photon change frequency, in case of the photon fail from a height H in a homogeneous gravitational field, with acceleration of gravity g.

Such situation was realized in well-known Pound and Rebka experiments [26]. Energy change corresponds to such frequency shift is equal

$$\Delta E = \left(\frac{\hbar\omega}{c^2}\right) gH. \qquad (32)$$

According to the formula (7) in the case of rotation in plane (TS) photon gains a mass

$$m = \frac{\hbar\omega}{c^2}\sqrt{n^2 - 1} \qquad (33)$$

The difference of potential energies in the point of emission and point of absorption of a photon, which differs by height H, is equal to

$$\delta E = mgH = \left(\frac{\hbar\omega}{c^2}\right) gH\sqrt{n^2 - 1}. \qquad (34)$$

Near to a surface of the Earth refractive index n of gravitational field is define by the formula (22). Taking into consideration an inequality (27), we will receive an evaluation

$$\delta E \approx \left(\frac{\hbar\omega}{c^2}\right) gH\sqrt{n^2-1} \approx \left(\frac{\hbar\omega}{c^2}\right) gH\sqrt{\frac{2\gamma M}{Rc^2}} = \left(\frac{\hbar\omega}{c^2}\right) gH\sqrt{\frac{2gR}{c^2}} \approx \left(\frac{\hbar\omega}{c^2}\right) gH(2.5 \cdot 10^{-5}). \qquad (35)$$

We can see, that correction to effect connected to photons nonzero mass emerging, near to the Earth surface is only $10^{-5}$ from magnitude of the total effect.

## 4.3 Radar echo delay

Let's assume that we measure time of light signal propagation up to some object, and back in space. Radar echo delay can differ in dependence from that, does this light spread in empty space, or in a gravitational field. Such delay was measured in experiments on location of Mercury and Venus [27]. The experiments give satisfactory agreement with General Theory of Relativity (GR) predictions.

These experiments also were analyzed in [28]. Here we do not interesting to analysis of these work. We want only to show that the analytical expression for magnitude of delay of a radar echo in ESM coincides what is received in GR. This result can be obtained from the fact that the photon time delay $\Delta t$ is calculated only from the photon velocity v(t) [24,25]. Let's imagine that we locate the Sun. We have

$$\Delta t = 2\left(\int_{R_s}^{r_e} \frac{dr}{v(r)} - \int_{R_s}^{r_e} \frac{dr}{c}\right). \quad (36)$$

Here $R_s$ is radius of the Sun, $r_g$ - gravitational radius of the Sun, and $r_e$ - distance from the Earth up to the Sun. Speed of light in a gravitational field is $v = c/n$. As to here we deal with photons, it is necessary to select refractive index $n = n_1$. Substituting it in (36), we obtain

$$\Delta t = \frac{4\gamma M}{r_e c^2} \cdot \ln\frac{r_e}{R_s}. \quad (36)$$

Formula (36) coincides with expression for magnitude of radar echo delay obtained in [25,28]. It is also possible to obtain in the frame of ESM the perihelion precession of Mercury and magnitude of light beam deviation angle from a rectilinear trajectory in case of photon propagation near to a massive body [3].

Thus in ESM framework it is possible to obtain the same predictions as in the GR.

GR predictions can be received in ESM based on analogies between optical and mechanical phenomena. For this purpose it is enough to use only technique of rotations in the extended space G(1, 4) and formula for the refractive index in the Schwarzschild metric. Any additional ideas and suppositions were not attracted. Actually obtained results are only first approximation in an evaluation of magnitude of gravitational effects.

## 5. 5-DIMENSIONAL ELECTRODYNAMICS IN EXTENDED SPACE MODEL

In [1,2] in the ESM were constructed: 5-vector potential –

$$(\varphi, \vec{A}, A_4) = (A_t, A_x, A_y, A_z, A_s), \quad (37)$$

stress 5x5 tensor -

$$F_{ik} = \frac{\partial A_i}{\partial x_k} - \frac{\partial A_k}{\partial x_i}; i, k = 0,1,2,3,4 \quad (38)$$

and energy-momentum-mass tensor -

$$T^{ik} = \frac{1}{4\pi}(-F^{il}F_l^k + \frac{1}{4}g^{il}F_{lm}F^{lm}); i,k,l,m = 0,1,2,3,4. \quad (39)$$

Here $g = \|g^{ik}\|$ is metric tensor of the extended space. The stress tensor record as:

$$\|F_{ik}\| = \begin{bmatrix} 0 & -E_x & -E_y & -E_z & -Q \\ E_x & 0 & -H_z & H_y & -G_x \\ E_y & H_z & 0 & -H_x & -G_y \\ E_z & -H_y & H_x & 0 & -G_z \\ Q & G_x & G_y & G_z & 0 \end{bmatrix} \quad (40)$$

where the elements with indexes i, k = 0, 1, 2, 3, 4 correspond to components of stress tensor of an electromagnetic field in 4-D space-time. Vector field $\vec{G} = (G_x, G_y, G_z)$ and scalar field Q are new additional components.

In the energy-momentum-mass tensor, which associated with density of energy, are added additional components $G^2$ and $Q^2$ also

$$T^{00} = \frac{1}{8\pi}(E^2 + H^2 + G^2 + Q^2) \quad (41)$$

Modified vector of the momentum density in ESM looks like as:

$$\vec{P} = \frac{1}{c}(T^{01} + T^{02} + T^{03}) = \frac{1}{4\pi c}([\vec{E}, \vec{H}] - Q \cdot \vec{G}) \quad (42)$$

By means of the stress tensor (38) and the vector of the 5-d current $\vec{\rho} = (\rho, \vec{j}, j_s)$, extended Maxwell equations have been constructed in ESM [1,2]

$$\text{div}\vec{H} = 0 \quad (43)$$

$$\text{rot}\vec{E} + \frac{1}{c}\frac{\partial \vec{H}}{\partial t} = 0 \quad (44)$$

$$\text{rot}\vec{G} + \frac{\partial \vec{H}}{\partial s} = 0 \quad (45)$$

$$\frac{\partial \vec{E}}{\partial s} + \frac{1}{c}\frac{\partial \vec{H}}{\partial t} + \text{grad}Q = 0 \quad (46)$$

$$\text{div}\vec{E} + \frac{\partial Q}{\partial s} = 4\pi\rho \quad (47)$$

$$\text{rot}\vec{H} - \frac{\partial \vec{G}}{\partial s} - \frac{1}{c}\frac{\partial \vec{E}}{\partial t} = \frac{4\pi}{c}\vec{j} \quad (48)$$

$$\text{div}\vec{G} + \frac{1}{c}\frac{\partial Q}{\partial t} = 4\pi j_s \quad (49)$$

Detail analysis of the extended Maxwell's system (43)-(49) look at [1].

## 6. MECHANISM OF THE DARK MATTER AND CONDENSED BUBBLE OBJECTS FORMATION IN THE EXTENDED SPACE MODEL

Within the framework of Extended Space Model the processes connected to birth of photons in a gravitational field are studied also. These photons have a nonzero mass. It can be both positive, and negative, and photon's energy and strength of the gravitational field determine its absolute value [3].

It was shown that in ESM model formation of bubble gravitational structures is possible.

In the frame of ESM one could obtain the follow physical picture. Bubble gravitational objects have a halo formed by dark matter generated by photons with a positive mass. The photons with a negative mass are throw away in free deep space and create there antigravitating vacuum with negative pressure [3].

In the ESM frameworks there is a mechanism, according to which in a neighborhood of massive gravitational objects the halo consisting of photons with positive masses can be formed. These halos we associate with a dark matter. The photons with negative masses are concentrated far from massive objects. These photons will create areas of a dark energy.

Such areas are characterized by negative pressure and exhibits properties of antigravitation. This area calls the accelerated extension of that visible part of the universe, which consists of a positive matter.

## CONCLUSION

In the present work the generalization of Einstein's special theory of relativity on 5-dimentional extended space is considered, in which as fifth coordinates we introduce the interval s of a particle. This generalization special theory of relativity on 5-dimensional space, allows constructing 5- vector p = (E/c; $p_x$, $p_y$, $p_z$, mc), where as fifth component the mass of a particle introduced.

All components 5-dimensional of a vector are connected by a well-known formula:

$E^2 = c^2 p_x^2 + c^2 p_y^2 + c^2 p_z^2 + m^2 c^4$. Within the framework of such approach there is no difference between massive and massless particles. In difference from 4-dimensional Minkowski space, in the entered extended space the 5-vectors become isotropic both for mass, and for massless particles.

As a result of magnification of dimensionality of space there are, except for the Lorentz transformation two another transformations of space (hyperbolic and Euclidean turns), which transfer massive particles into massless and visa versa.

During all of these transformations isotropy of 5-vectors are not lost.